# Implications for Improving Accessibility to E-Commerce Websites in Developing Countries - A Subjective Study of Sri Lankan Hotel Websites

Arunasalam Sambhanthan[1], Alice Good[2]
School of Computing, University of Portsmouth,
United Kingdom
arunsambhanthan@gmail.com[1], alice.good@port.ac.uk[2]

## Abstract

This research explores the accessibility issues with regard to the e-commerce websites in developing countries, through a subjective study of Sri Lankan hotel websites. A web survey and a web content analysis were conducted as the methods to elicit data on web accessibility. Factors preventing accessibility were hypothesized as an initial experiment. Hazardous design elements are identified through web content analysis, the results of which are utilized to develop specific implications for improving web accessibility. The hypothesis tests show that there is no significant correlation between accessibility and geographical or economic factors. However, physical impairments of users have a considerable influence on the accessibility. Especially, visual and mobility impaired users experience poor accessibility. Poor readability and less navigable page designs are two observable issues, which pose threats to accessibility. The lack of conformance to W3C accessibility guidelines and the poor design process are the specific shortcomings which reduce the overall accessibility. Guidelines aim to improve the accessibility of sites with a strategic focus. Further enhancements are suggested with adherence to principles and user – centered design and developing customizable web portals compatible for connections with differing speeds. A need for developing new design models for differencing user groups and implementing web accessibility strategy are emphasized as vital steps towards effective information dissemination via e-commerce websites in the developing countries.

## 1. Introduction

Accessibility is a critical factor for e-commerce success. Particularly, the accessibility of e-commerce websites for users with impairments is an evolving research area in the contemporary academia. Apart, the accessibility of e-commerce websites in developing countries could be influenced by several other factors such as economic background and geographical location of users. Therefore, it is a vital requirement for web concept designers in the developing countries to consider the above factors in order to ensure the successful dissemination of business information through e-commerce sites. However, the means of designing accessible web interfaces is still remaining as an open question with regard to developing countries.

This study aims to formulate implications for improving accessibility to e-commerce websites in developing countries through a subjective study of Sri Lankan hotel websites. Specially, formulating implications for web designers in developing countries would be a vital contribution to the body of knowledge. In addition to this, the specificity of this research has been focused to Sri Lankan hotel websites considering the contextual relevance. Arguably, Sri Lankan hotel websites have a greater need for accessibility due to the reservations made by tourists from different physical, economic and geographic backgrounds. The current economic boom, opened by the post war scenario of the island further signifies the emerging need for designing accessible e-commerce sites, for effective web based promotion.

## 2. Literature Review

According to Good (2008, p. 16) "A website is said to be accessible when anyone, regardless of economic, geographic or physical circumstances, is able to access it". Hence, for this particular study, accessibility could be defined as the ease in which people with disabilities, people from different geographic regions and people having different internet connections could access the websites'.

The economical aspect of accessibility could be looked in two ways, (1) the category of internet connection - which depends on the economic background and usage frequency of customer, and (2) the time spent on information search which mostly depends on the time availability of consumer. Both these factors have an interrelationship and will have a direct effect on accessibility. The type of internet connection determines the speed of browsing, which influences the time a consumer spends on browsing. The less speedy the connection, the more the consumer will get discouraged and vice-versa. On the other hand consumer's average browsing time per day determines the category of connection they hold in the context of a developing country. However, search engine optimization - an evolving area in tourism research deals with this issue to a certain extent. The degree of filtration and the sequence of listings done by search engines will influence the accessibility of tourism sites. Law & Haung (2006) indicates that 47.4% users only check the first three pages of the search results listing. The above finding has not specified the number of results listed per page. However a recent study of Xiang *et al* (2008) moderates the previous finding, stating that travelers are only retrieving a tiny fraction of the huge list of search results. Evidently, search result listing is a dominant factor which could influence the decision behavior of consumers. But, it could be argued whether the curious consumer will not spend more time to get the best out of his search. Also, the above rule might be inapplicable for frequent customers as they would directly go to the hotel web page to check the information. However the information search behavior of consumers will trigger to have a look for 'innovation' in each hotel's website. On the other hand, hotels will not be able to survive in today's knowledge economy with their 'obsolete' ways of promoting products. Consequently, being in the forefront of search results listing and using a customer centric approach in wording websites are vital for getting the customers to stay with the site.

The geographic aspect is another major factor which determines accessibility. Accessing the website from rural areas with poor bandwidth is another challenge. Utilization of mobile commerce is another evolving area in tourism to solve this issue. According to Hyun *et al* (2009, p. 158) it is critical for hotels to utilize mobile commerce in tourism promotion to survive in an era of increasing mobile adaptation. In addition Buhalis (2008, p. 612) claims mobile technology increases the accessibility of digitally excluded communities. However, investing on mobile technology could pose a number of challenges to hotels. Pizam (2009, p. 301) indicates that the current global financial crisis has adversely affected the hospitality industry and tends the income of hotels to be reduced over time. On the other hand, Buick (2003, p. 244) identified customer demand as a pulling factor in IS/IT investment. Will it be advisable to invest on mobile technology in the light of the above contradicting dilemma? But the competitiveness gained through new technology could be powerful than any of the above factors. Supportively Kozak *et al* (2005, p. 7) ascertain that the ability to gain IT access is the deciding factor of first and second class tourist destinations, organizations and customer groups.

Finally, the influence of physical circumstances has a major role in accessibility. Eichhorn *et al* (2008) argues more sophisticated understandings of differential needs and appropriate sources as the most crucial step in enabling disable tourists to access tourism information. But this statement could be discarded, as there are almost enough specific needs identified by disability scholars, and the labor is entirely on the hands of technology researchers to contextualize and develop design frameworks for users with impairments. On the contrary, the doctoral research of Good (2008) investigated a method to improve the web accessibility of users with impairments. However; the profundity of Good's (2008) work could be criticized for its limitation to cover the whole population of disable consumers. Also the sample selected by Good (2008) is very small and could be claimed as inadequate to generalize the findings. However; Good (2008, p. 134) defended her research claiming that excellent results could be achieved using small groups. It is acceptable, as all the visually impaired users will face difficulties in reading the letters with small font size. Anyhow, further research is needed to extend the findings of Good (2008) and to contextualize which for the tourism domain. Concluding from the above findings, interface usability is the dominant factor deciding the web accessibility of people with impairments.

### 3. Methodology

A user centric approach was adapted for this research. The data collected through a web survey and web content analysis. Initial hypothesis testing was conducted based on the data collected through web survey. The hazardous design features which prevent accessibility were triggered through the web content analysis. The feasibility of collecting data from geographically dispersed samples was the main reason for selecting a web based survey for data collection. The web content analysis selected for specific identification of design shortcomings.

A website was created to provide the preliminary information to survey respondents. The web based survey was piloted with 7 participants representing all three sample groups to improve the usability of it. The improvement of language, inclusion of navigational links and the incorporation of rationale for questions were the notable outcomes of pilot study. The hotels falling under western region of Sri Lanka (22) were selected and the hotels with e-commerce websites were shortlisted based on a consequent Google search.

The number of questionnaires to be promoted was processed as 120. In fact the maximum number of dependent variables was determined as 4. Altogether 80 responses were estimated deciding 20 responses per variable. To reach 80 responses it was decided to gather 120 responses considering 40 invalid responses. Considering the above outcome the survey was planned for one full week. Links of all 22 hotel sites were given to the respondents and they were asked to browse one website on their own choice and to record their experience.

The elements rated by users analyzed through a quantitative approach. The analysis of statistical variance was undertaken to check the validity of each sub hypotheses. ANOVA test of user ratings was undertaken using online analysis software. The web content of selected five hotels was analyzed against a checklist prepared based on a randomly selected criterion from W3C web accessibility guidelines. Apart from this, short interviews were conducted with hotel managers to explore the organizational initiatives with regard to improving accessibility.

### 3.1. Sample Groups

The study was conducted in the natural environment of user. 120 participants from 14 countries participated in the survey. The sample was classified into different groups check the accessibility of websites for users from varied economic, physical and geographic backgrounds. User samples included connection types [ADSL broadband and any other], impairments categories [visual, mobility and cognitive / language impairments] and geographic regions [Sri Lanka and foreign users].

The connection types were classified into three main categories namely ADSL, broadband and any other connections. Any other connection category includes dial up connection as well. The first two connections represent people having good bandwidth and speedy internet accesses, while the other category denoted people with below average bandwidth.

The geographic destinations have divided into two categories namely, Sri Lankan and foreign users. Rationale for which was to assess the accessibility for both within and outside Sri Lanka. Justifiably, the research not intends to assess the country wise accessibility of sites, instead accessibility for different geographic destinations.

Special user needs have classified into four main categories namely, visual, auditory, cognitive/language and mobility. The categorization of visual and auditory needs includes all the impairments related to vision and hearing. The cognitive/language needs include impairments related to language, speech and cognition. The mobility includes dexterity and motor impairments. Good (2008, pp. 28) classified impairments into the above categories to demonstrate the relationship between design features and accessibility.

The above classification was adopted as it gives a comprehensive method for classifying disabilities with regard to internet accessibility. Arguably, one of the objectives of this study is to measure the accessibility of sites for users with different impairments, but not to measure the internet accessibility for individual impairments. However, the auditory impairment was eliminated from the list in later stages as there were not enough responses recorded from participants representing that group. A brief description of the above categorization was given in the research website and linked with survey to enable respondents understand the method followed for classification.

While there may be some criticism on selecting a sample of generic users instead of tourists, it is defendable in this instance. The purpose of study is to measure the accessibility of sites. Accessibility of a site is largely depending on aspects such as users' geographic, economic and physical backgrounds and site design (Good 2008). In fact, the accessibility of a site would be same for a tourist and non tourist user.

## 4. Results

This section presents the accessibility ratings of users. Participants asked to rate the accessibility of sites. The correlation between accessibility and living zone, type of connection and physical impairments were measured accordingly. The, user ratings were aggregated and presented as a table. (SD: Strongly Disagree, D: Disagree, NAND: Neither Agree nor Disagree, A: Agree, SA: Strongly Agree)

| Classification | SD | D | NAND | A | SA |
|---|---|---|---|---|---|
| Sri Lanka | 1 | 1 | 9 | 20 | 3 |
| Other Countries | 0 | 1 | 22 | 19 | 4 |
| Cognitive / Language | 0 | 0 | 2 | 2 | 0 |
| Mobility | 0 | 1 | 1 | 2 | 0 |
| Visual / Sight | 0 | 1 | 0 | 2 | 1 |
| ADSL | 0 | 0 | 15 | 18 | 4 |
| Broad Band | 1 | 2 | 15 | 19 | 3 |
| Dial Up | 0 | 0 | 1 | 1 | 0 |
| Any Other | 0 | 0 | 0 | 1 | 0 |

**Table 1: Accessibility Ratings of Users**

Country wise accessibility scores show most of the responses are above 'neither agree nor disagree' category. There are no users with auditory impairments. Most of the users with cognitive/language and mobility impairments have rated above average score for accessibility. However a visually impaired user disagrees with the accessibility. Most of the responses for connection wise accessibility show above average category. Thus, generally the websites could be said to be accessible for users from varied geographic, economic and physical backgrounds, except the tiny fraction of negativity recorded from users with visual and mobility impairments. However, the statistical mean and median for each rating were calculated for further analysis.

## 5. Statistical Analysis

Table 2 presents the statistical ANOVA results of the connection wise accessibility scores.

| Source of variation | Sum of squares | d.f. | Mean squares | F |
|---|---|---|---|---|
| Between | 0.3574 | 2 | 0.1787 | 0.391 |
| Error | 30.63 | 67 | 0.4571 | |
| Total | 30.99 | 69 | | |

**Table 2: Statistical Analysis of Connection wise Accessibility Scores**

The probability of this result assuming null hypothesis is 0.68. Therefore the null hypothesis is valid and the research hypothesis cannot be accepted. Hence, the type of connection does not have any significant influence over the accessibility of sample websites. Therefore, the economic background of the user does not need to be a deciding factor of accessibility according to the above hypotheses.

Table 3 depicts the t- test results for the region wise accessibility scores of websites.

| Region | Mean | Standard Deviation | Median |
|---|---|---|---|
| Sri Lanka | 3.66 | 0.614 | 4 |
| Foreign Countries | 3.54 | 0.711 | 3 |

**Table 3: Statistical Analysis of Country wise Accessibility Scores**

Especially, the regions have been divided into two main groups as Sri Lanka and foreign countries, and a hypothesis was tested to scrutinize whether the geographic diversity has any significant effect on the accessibility of websites. The results indicate that the probability of these results assuming the null hypothesis is 0.47, which is lesser but much closer to the middle point. However, the mean and median of the accessibility scores shows a more than average score for accessibility from differing regions. Especially, the raw data reveals that almost 90% users rated 'above average' for accessibility. Hence, it is concluded that the null hypothesis is valid and there is no significant influence on accessibility by the regional diversity.

Table 4 depicts the statistical ANOVA results of the website accessibility scores of users with special needs.

| Source of variation | Sum of squares | d.f. | Mean squares | F |
|---|---|---|---|---|
| Between | 0.5 | 2 | 0.25 | 0.2647 |
| Error | 8.5 | 9 | 0.9444 | |
| Total | 9 | 11 | | |

**Table 4: Statistical Analysis of Accessibility Scores of Users with Special Needs**

The probability of these results assuming the null hypothesis is 0.77. Therefore the null hypothesis is valid and the research hypothesis cannot be accepted. Thus, the website accessibility cannot be proved to be influenced by the individual impairment. However, the raw data depicted in table 1 reveals, one each individual users from visual / sight and mobility impairment groups disagree with the accessibility of sites. A site inaccessible for a user with sight disorder, due to the lack of readable fonts or alter text features, should be inaccessible for all users with the same level of sight disorder. Hence, the sites are inaccessible for users with visual and mobility impairments. Abanumy *et al* (2005) reports the same phenomenon among the e-government websites of Saudi Arabia and Oman. Therefore this could be generalized as a general phenomenon in website design. The following section formulates the accessibility guidelines for hotel industry website designers. Thus, these results should be subjected to further analysis to identify the exact design elements which prevent the accessibility of sites for users with visual and mobility impairments. Consequently, a web content analysis of five sample sites was conducted based on randomly selected criterions from W3C accessibility guidelines. The shortcomings identified with the web content have been used extending the findings of preliminary survey and discussed in the proceeding section.

## 6. Design Elements

One each guideline from all four design principles, were randomly selected from the four main categories of W3C guidelines for analysis. The selected criterions are: (1) Text alternatives, (2) Navigable, (3) Readable, (4) Compatible. More than 60% of the sites reviewed do not have alt-text facility. This will have a significant effect for users with visual impairments. The navigation through pages is observed to be taking more time and this will definitely de-motivate the user. Especially, the location specification of the web pages has not been indicated during the course of navigating through pages. A very poor readability is observed in more than 60% of the sites analyzed. Especially, the sites do not have fonts which are visible to the users with sight impairments, which is mostly because of the inappropriate usage of color combination in the font selection. One hotel uses photo enlarging option to support visually impaired users. Another hotel holds two options in their homepage. One is for customers with slow connections and other one is for speed connections. Basic product information and company information are presented in both sites. Although, the site compatible with slow connections have designed with less multimedia objects, which lacks to incorporate readable fonts in web design.

## 7. Implications

- > *Adherence to principles:* The web content analysis showed a general weakness of websites with regard to lack of adherence to W3C accessibility guidelines. Disregarding W3C guidelines had led to poor design, which reduces the readability of websites to people with visual impairments. Also, the mismatch with W3C guidelines has prevented sites from being accessible to mobility users. This could be synthesized as a common weakness in most of the commercial websites from developing regions.

- > *User-Centered Design:* There are certain arguments on the danger of solely depending on the accessibility

guidelines (Milne *et al*, 2005). On the other hand, the exact need of the user could be triggered solely by including user in design process. Although, this practice is well expounded in academic researches, it is identified as a common weakness in hotel websites. Therefore, it is recommended to adapt a user - centered approach in web design. Especially, it is vital to include users with visual and mobility impairments as well some elderly users in the design process, to ensure the responsiveness of websites to the needs diverse user groups.

➢ *Design Frameworks*: Specific design models could be developed to support web designers in the developing countries. The development of design models and testing the design with end users would immensely increase the accessibility of websites. Although, the above best practices are well practiced by most of the web designers in the developed world, a major setback on accessibility concerns has been observed among the organizations located in the developing countries.

➢ *Portals with different speeds*: Although the hypothesis tests shows there is no co-relation between the accessibility and the type of internet connections, the results of this study has been obtained from a comparatively small sample for connection wise users groups. Especially, the sample for users with other connection types was comparatively small. On the other hand, there is a high probability for greater number of users from developing countries holding low speed connections. Hence, it is recommended to design websites with customized portals to match with the differential speed categories of users.

➢ *Accessibility Strategy*: A lack of accessibility strategy among the hoteliers has been clearly scrutinized through the above research. Especially, the websites are not having any proper standard with regards to the accessibility for users with differing needs. This could be scrutinized through improper color combination and lack of clarity of fonts. Particularly, the sites are designed on the convenience of designers than of users. The lack of mutual contribution between academia and industry and poor awareness level on accessibility issues could be the main reasons for the above phenomenon in developing countries. Christian (2009) reported a similar phenomenon with Australian small hotels and suggested to adapt a web strategy to avoid lapse. However, an accessibility strategy would be a sensible solution for the above issue in the context of this research. As part of accessibility strategy, the organizations could organize knowledge transfer sessions for web designers and walkthroughs with users. This will help the designers to grasp a high level understanding of the accessibility issues as well as to elicit the exact user requirements with regard to accessibility.

## 8. Evaluation & Conclusions

The study investigated on implications for improving the accessibility of e-commerce websites in developing countries. The study compiled with a case study of Sri Lankan hotel websites. The results indicate poor accessibility for users with visual and mobility impairments. Poor readability and less navigable pages in the websites were scrutinized as the principle reasons which prevent accessibility of sites for users with visual and mobility impairments. Also, the sites have been observed not adhering to W3C accessibility principles. The implications includes adherence to principles, user centered design, developing design frameworks, portals with different speeds and an accessibility strategy.

The study lays a foundation for web accessibility for websites research in developing countries. Especially, this is an explorative study for Sri Lankan hotel industry and the results will shed light on future research avenues towards developing specific design models for accessible web design. Also, the outcome could be extended further by developing testable prototypes to address differencing user needs, testing which with different user groups would reveal specific contextual needs of tourists with special needs.

On the other hand, this study could be criticized for a number of limitations. Firstly, the research hypothesis was developed solely based on a predisposition that the economic diversity of the sample could be exactly maintained through including users with different connection types. It could be argued that there could be other different parameters which could form a clear foundation for accessibility research of users from different economic backgrounds. Secondly, the selection of sample consists of generic users instead of tourists. However, it is defendable as the accessibility is same for a tourist and non-tourist. In addition to this, the non equal distribution of samples among the sample groups could be argued as another inherent weakness of this research. Although, accessibility for elderly people is an evolving area (Good *et al*, 2007), they have not been included in the study. In addition to this, the web content analysis is an unstructured one and could be improved by doing a detail analysis of the web content against W3C guidelines. That would results in developing new design frameworks for e-commerce sites in the developing countries.

In conclusion this research is just a beginning for Sri Lankan hotel industry. Further research needs to be undertaken to extend the findings on factors influencing the accessibility of sites for users from different economic, geographic and physical backgrounds. However, this particular study remains as beginning for Sri Lankan hotel industry, while adding some facts to the existing knowledge pool of accessibility research literature for the context of developing countries.